\documentclass{article}
\usepackage{spconf,amsmath,graphicx,hyperref,booktabs,subcaption}
\usepackage{comment}
\title{MTCRNN: A multi-scale RNN for directed audio texture synthesis}
\name{Muhammad Huzaifah \hspace{0.8cm} Lonce Wyse\thanks{This research is supported by a Singapore MOE Tier 2 grant ``Learning Generative Recurrent Neural Networks'' and by an NVIDIA Corporation Academic Programs GPU grant.}}
\address{National University of Singapore, Singapore \\ \normalsize E0029863@u.nus.edu, lonce.wyse@nus.edu.sg}
\begin{document}
\ninept

\maketitle
\begin{abstract}
Audio textures are a subset of environmental sounds, often defined as having stable statistical characteristics within an adequately large window of time but may be unstructured locally. They include common everyday sounds such as from rain, wind, and engines. Given that these complex sounds contain patterns on multiple timescales, they are a challenge to model with traditional methods. We introduce a novel modelling approach for textures, combining recurrent neural networks trained at different levels of abstraction with a conditioning strategy that allows for user-directed synthesis. We demonstrate the model's performance on a variety of datasets, examine its performance on various metrics, and discuss some potential applications.

\end{abstract}
\section{Introduction}
\label{sec:intro}

Audio ``textures'' have proven elusive to define precisely despite their ubiquity and long history of attention in auditory and modelling research. Saint-Arnaud and Popat \cite{saint1995analysis} described them as having stable characteristics over the time scale of an ``attention span''. That is, there is a duration beyond which an audio texture provides no new information. Many common environmental sounds, both natural and human generated, seem to have this characteristic, including the sound of flowing water, a ticking clock, running machines, crackling fire, or a rumbling engine. The working definition suggests that for durations larger than some attention span, any two analysis windows at different points in time will reveal the same properties.

How then do we contextualise sounds that seem ``textural'' but might also exhibit changes over the course of time, such as the sound of an accelerating car or intensifying rain? One way is to recognise that what we deem as the stable textural part of a sound may exist as an abstraction (a perceptual or computational model) in tandem with another layer that is modulating the overall sound \cite{aicreative}. While the former is representative of the essential and unchanging audio characteristics that deﬁne a texture class, the latter includes the aspects that modify parts of the sound but do not alter the overall perceptual identity of the texture. Capturing these factors of variation -- imagine an engine revving at two different speeds and the resultant sound differences -- and exposing a way to navigate the space encompassed by them, are arguably essential to a complete model of a texture.  


Here we describe and develop the multi-tier conditioning recurrent neural network (MTCRNN) that embodies the multi-level paradigm of the constant fine structure and the evolution of the texture over time. The system aims to synthesise textures directly in the raw audio domain guided by user-determined control parameters. The challenge of capturing both local and long-term dependencies present in raw audio is met with a combination of hierarchically-arranged model architectures and multi-scale conditioning. More concretely, we condition the audio output on synthetically generated intermediate parameters based upon interpretable physical or perceptual audio features, which are in turn informed by more compact control parameters corresponding to an arbitrary path through a perceptual dimension of the texture of interest.





\section{Background}
\label{sec:background}

\subsection{Hierarchical deep generative models}
\label{ssec:hierarchy}
State-of-the-art generative models for raw audio have been based around the idea of hierarchically arranged layers to capture increasingly long time dependencies in the data. Wavenet \cite{oord2016wavenet} achieved this by using dilated convolutional neural network (CNN) layers where each convolutional kernel takes every $n$th element from the previous layer as input instead of a contiguous section. The RNN layers in SampleRNN \cite{mehri2016samplernn} are organised into tiers that learn at different levels of abstraction according to a particular timescale. Except for the lowest-level tier, all other tiers operate at a frequency below the sample frequency, each summarising the history of its inputs into a conditioning vector for the subsequent tier.

Even with these innovations to extend the receptive field of the models, their outputs were still found to lack coherent long-term structure without a strong external conditioning signal. More recent work has further decoupled the learning of local and larger-scale structures by training separate models at each level of abstraction. The latent space of the model working at longer timescales is used to condition another model operating at shorter scales. This was demonstrated by Dieleman et al. \cite{dieleman2018challenge} in an autoencoder setup with Wavenet decoders. A similar idea was implemented in OpenAI’s Jukebox system \cite{Dhariwal2020}, which utilises autoregressive transformers to model encodings at multiple timescales over time.

\subsection{Inducing long-term structure and control}
\label{ssec:struc}
A drawback to using latent representations is that they are often not interpretable and thus not easily manipulatable by the user. Alternatively, using conditioning from symbolic representations would, in addition to being more interpretable, be easier to model over the long term, while still preserving the richness and expressivity of the raw audio models themselves. In the musical domain, Manzelli et al. \cite{manzelli2018conditioning} provided MIDI pitch and timing as conditioning for a Wavenet. Rather than feeding these values by hand, a second generative network was used to compose MIDI sequences, taking over the modelling of long-range timing and pitch correlations. This idea was further expanded by Hawthorne et al. \cite{hawthorne2018enabling} in their Wave2Midi2Wave system, which added an encoder network to transcribe raw audio to MIDI and replaced the RNN generating MIDI sequences with a state-of-the-art transformer network. Compared to the hierarchical Wavenets in \cite{dieleman2018challenge} which learnt to interpret and generate intermediate latent representations, using MIDI yielded better samples in terms of musical structure, but were limited to the instruments and styles that MIDI could accurately represent. 

In our previous work \cite{aicreative}, we proposed the MTCRNN as a solution to generating long-form textures, and showed its advantages over an architecture that was trained only over a single timescale. Here, we apply the model to a larger variety of more complex textures and provide an extended analysis and evaluation of results. We additionally detail how the system may be utilised for style transfer.   


\section{Model}
\label{sec:method}

\subsection{Overview}
\begin{figure}[htb]
\begin{minipage}[b]{1.0\linewidth}
  \centering
  \centerline{\includegraphics[width=0.999\textwidth]{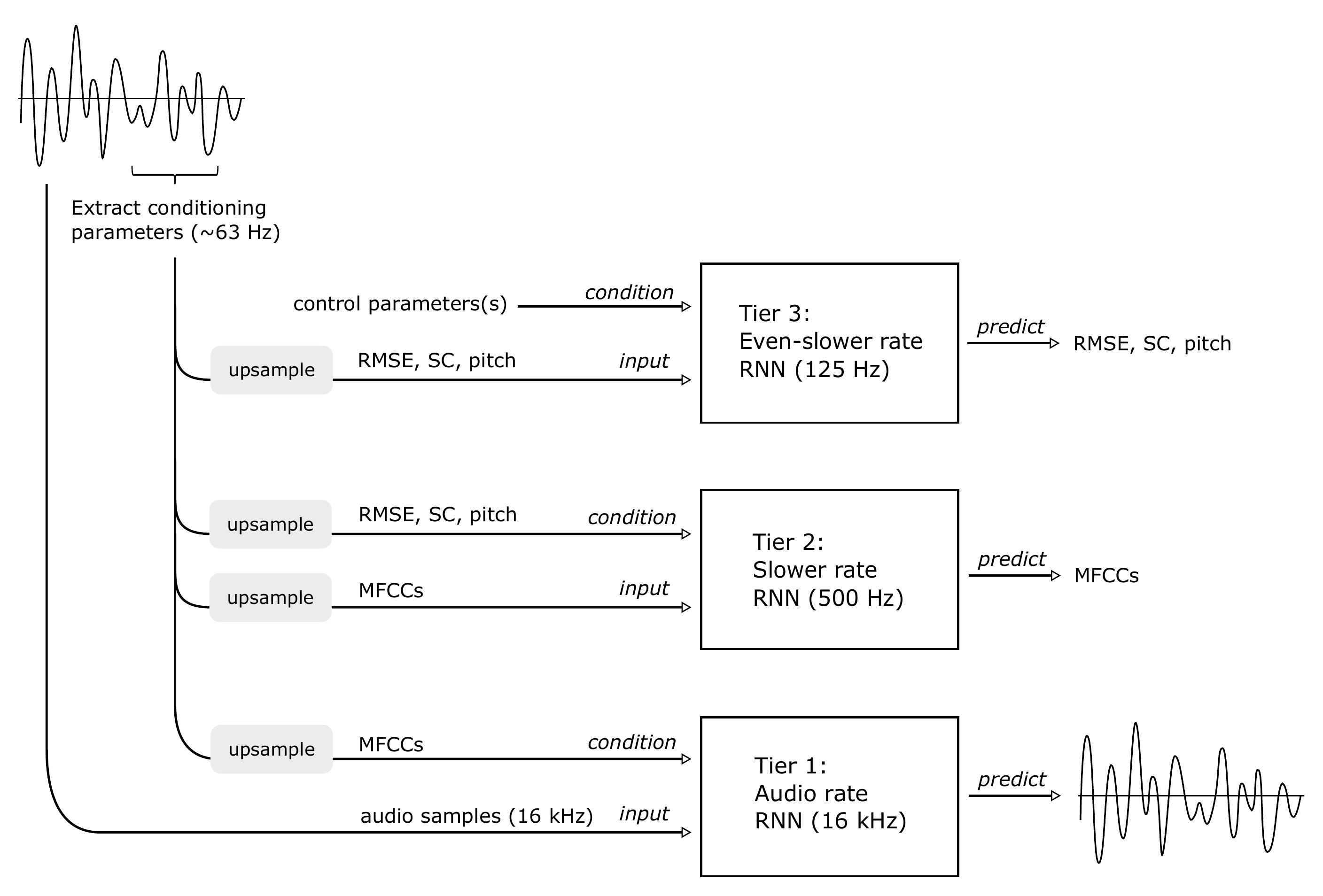}}
\end{minipage}
\caption{Example of training an MTCRNN with 3 tiers. Audio is used to train tier 1, while intermediate representations at lower sample rates are used to train other tiers. The user controlled parameter conditions the highest tier only.}
\label{fig:mtcrnn_train}
\end{figure}

The MTCRNN\footnote{\scriptsize \url{https://github.com/muhdhuz/MTCRNN}} combines an arbitrary number of smaller separately trained tiers made up of stacked gated recurrent unit (GRU) layers, with each tier operating at a particular timescale (Fig.\ref{fig:mtcrnn_train}). Every tier works autoregressively, generating an output conditioned upon the output of the previous timestep and auxiliary information from the preceding tier. The tiers are arranged in order of sampling rate, such that a tier working at a slower sampling rate conditions a tier handling faster changing data, down to the audio sample level. The key idea behind the model is to sequentially synthesise audio samples, while directed by conditioning parameters that capture patterns and the overall sound trajectory on a scale larger than sample level.

The tiered structure is proposed to address several issues. Firstly, like prior works, we hypothesise that it is more efficient for a model to focus on capturing patterns at a specific timescale rather than at many different ones simultaneously. Secondly, inputs get spaced out further in time moving up the hierarchy of tiers, extending the effective receptive field of the model given the same amount of data. Hence, it is possible to expose the model to longer duration signals at higher tiers. Although the conditioning signal consequently becomes increasingly sparse, the results show that this is still an effective trade-off. Finally, even though the model allows for the generation of intermediate conditional parameters with clear physical or perceptual meaning, some of these, like mel-frequency cepstral coefficients (MFCCs) and root mean square energy (RMSE), may not be immediately intuitive to the casual user in terms of its direct effect on the output sound, and furthermore require the manipulation of many parameters simultaneously. Breaking the model into tiers allows these parameters to be interfaced with more compact and easily manipulatable control variables while maintaining a rich enough conditioning layer to support the synthesis.

\subsection{Architecture}
Each tier is composed of stacked GRUs sandwiched between dense layers connected to the input and output, similar to the architecture in Wyse and Huzaifah \cite{wyse2019smc}. Preliminary experiments showed that wide hidden layers improved the quality of synthesis. Each stack is comprised of three GRU layers, each containing between 300 to 800 hidden units depending on the dataset and tier. Greater numbers than these were found to exhibit diminishing returns.

\subsection{Synthesis and control strategy}
\begin{figure}[htb]
\begin{minipage}[b]{1.0\linewidth}
  \centering
  \centerline{\includegraphics[width=0.999\textwidth]{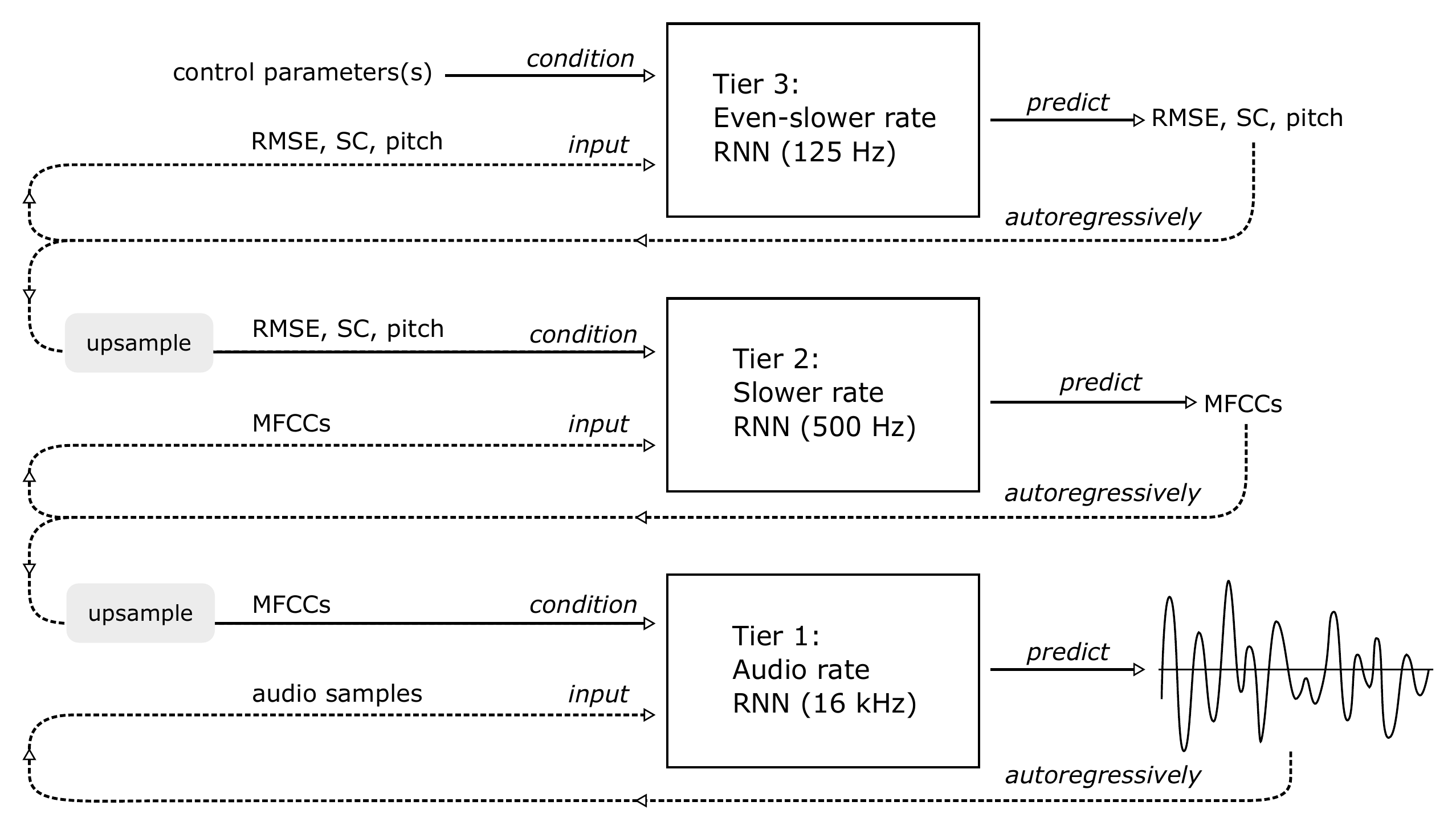}}
\end{minipage}
\caption{The full generative procedure for a 3-tier MTCRNN. The output sequence for the topmost tier is generated conditioned on the control parameters provided by the user, then upsampled to match tier 2’s sampling rate (i.e. 125 Hz to 500 Hz) and fed as conditioning to the tier 2 model. The process repeats for the lower tiers. In this example, MFCCs are used to condition the audio model itself.}
\label{fig:mtcrnn_gen}
\end{figure}

In the standard setup, user-driven control parameters are provided as input to the topmost tier. The model then generates an intermediate sequence using the control parameters as conditioning. Analogous to how MIDI was used for music in \cite{manzelli2018conditioning} and \cite{hawthorne2018enabling}, the intermediate sequences correspond to well-known audio features (not latent), but are real-valued as opposed to being discrete. Depending on the dataset, different combinations of features were used, including the aforementioned MFCCs and RMSE, but also pitch, spectral centroid (SC), and onset strength (Table \ref{table:c7_data}). These features have proven to be robust and applicable to a wide variety of sounds from prior literature. Conditioning for subsequent tiers are generated in a cascading fashion, and are upsampled by linear interpolation to match the sample rate of the corresponding tier prior to input, as shown in Fig.\ref{fig:mtcrnn_gen}.

Sampling from the audio model utilises a procedure similar to Wavenet's, whereby the categorical distribution described by the output of the network is sampled randomly to retrieve a $\mu$-law value for the sample at that timestep, then decoded for audio. In contrast, generated parameters are modelled as a Gaussian distribution, with the model treated as a mixture density network \cite{bishop1994mixture} predicting an associated mean and variance at each step.


Sound sets used for this study are from the Parameterized Audio Textures Data Sets (PATSets)\footnote{\scriptsize \url{https://sonicthings.org:9999/}} \cite{aicreative} along with heartbeat sounds from the PASCAL CHSC2011 dataset \cite{pascal2011}. A detailed breakdown of the various datasets and the corresponding parameters is shown in Table \ref{table:c7_data}. One MTCRNN model was trained per texture class.

\begin{table}[tb]
\centering
\scriptsize
\begin{tabular}{@{}lllll@{}}
	\toprule
	\multicolumn{1}{l}{\textbf{Texture dataset}}  & \textbf{Control pm.} & \multicolumn{2}{l}{\textbf{Conditioning pm.}} \\
	\cmidrule(lr){1-1}  \cmidrule(lr){2-2} \cmidrule(lr){3-4}
	&  &   \textbf{Tier 3} & \textbf{Tier 2}\\
	\midrule
	container filling & fill height & RMSE, SC, pitch & MFCCs  \\
	{[\url{https://bit.ly/3iZIgBv}]} & & & \\
	\midrule
	heartbeat  & class, rate  & -- & RMSE, SC, MFCCs \\
	{[\url{https://bit.ly/2SZz2KR}]} & & & \\
	\midrule
	engine  &  rev & -- & RMSE, pitch, MFCCs  \\
	{[\url{https://bit.ly/3k5M2dQ}]} & & & \\
	\midrule
	fire  & crackle  & -- & RMSE, MFCCs \\
	{[\url{https://bit.ly/37fYaFs}]} & & & \\
	\midrule
	geiger counter  & rate & -- & RMSE, onset strength\\
	{[\url{https://bit.ly/3j2Ifg8}]} & & & \\
	\midrule
	pop  & rate & -- & RMSE, onset strength\\
	{[\url{https://bit.ly/2SXzJ7x}]} & & & \\
	\bottomrule
\end{tabular}
\caption{Dataset properties, including the respective user control parameters and the conditioning parameters for each tier. Either three or two tiers were used for each dataset. Each texture class had control parameters related to changes in one or more high-level properties. The first two datasets are natural, while the rest are synthetic. Two versions of the ``pop'' texture was used, with higher rates in one than the other. A URL link to the results is provided for each dataset.}
\label{table:c7_data}
\end{table}


\section{Results}
\label{sec:exp}

\subsection{Frequency reproduction}
\begin{figure}[htb]
\begin{minipage}[b]{1.0\linewidth}
  \centering
  \centerline{\includegraphics[width=0.93\textwidth]{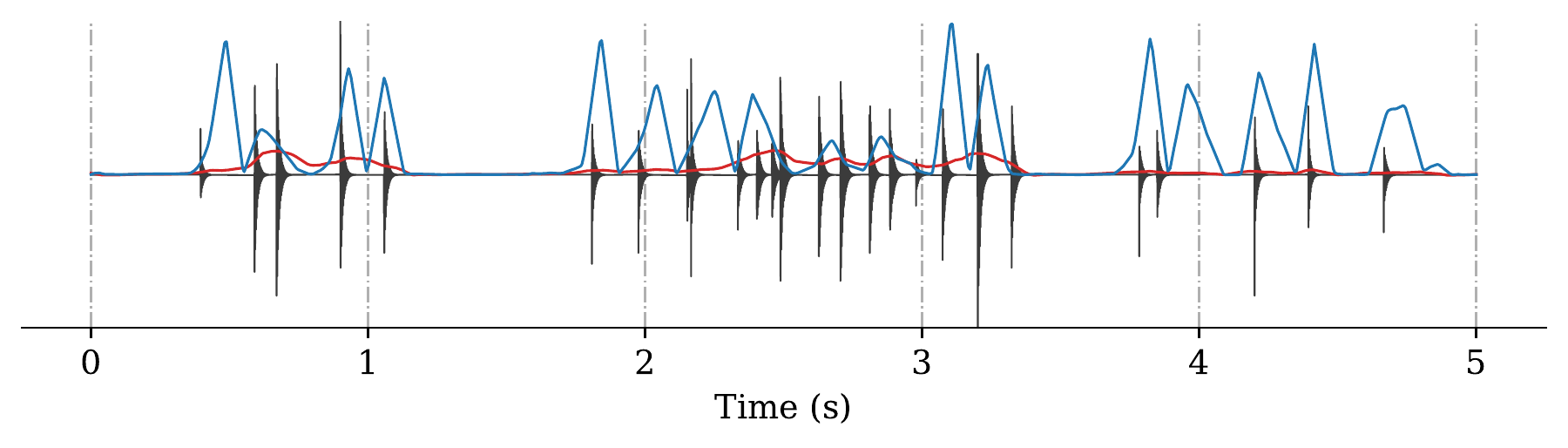}}
\end{minipage}
\caption{Synthesised ``geiger counter'' texture showing the close alignment between generated onset strength (blue) and RMSE (red) conditioning parameters, and audio (black) over long timescales.}
\label{fig:align}
\end{figure}

\begin{figure}[htb]
\begin{minipage}[b]{1.0\linewidth}
  \centering
  \centerline{\includegraphics[width=0.93\textwidth]{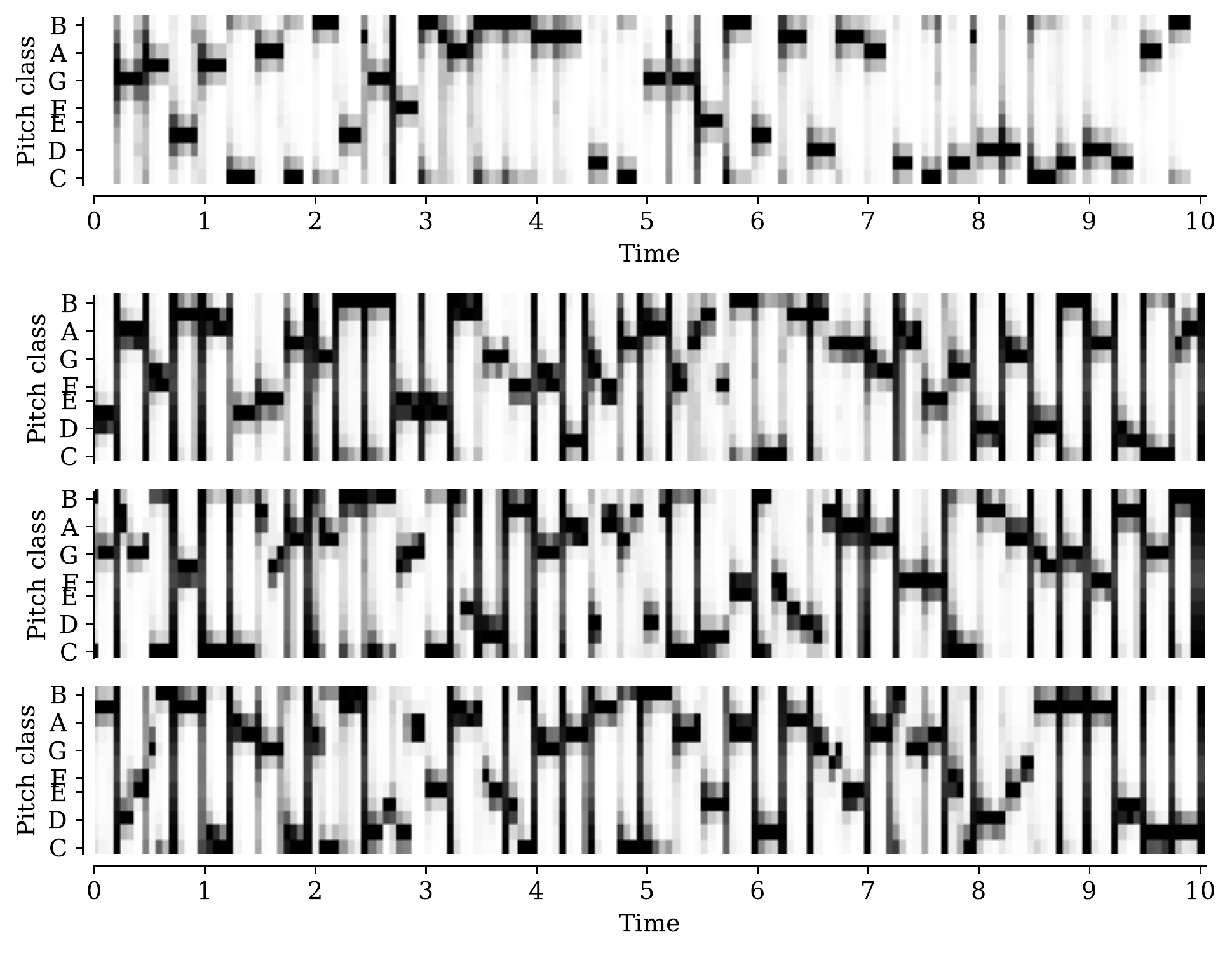}}
\end{minipage}
\caption{Chromagrams of the ``pop'' texture from the real dataset (top) and three separate generated instances (bottom), each individually unique. A darker colour indicates greater energy in that chroma bin.}
\label{fig:randpop_chroma}
\end{figure}

Long time dependencies can be addressed by a multi-tier architecture \cite{aicreative} as shown by Fig.\ref{fig:align}, while maintaining performance on local structure such as frequency patterns. The model was able to learn properties of both pitched and stochastic textures. A chromagram of the ``pop'' texture, consisting of noise bursts with narrow bandpass filters heard as pitched sounds, was plotted for both original and synthesised data (Fig.\ref{fig:randpop_chroma}). The chromagram reveals the fairly definite pitch class of each synthesised pop, resembling the real data. Moreover, the pitches are randomly distributed across the entire scale over the full duration of the sound file, a clear indication that the model has learned the inherent variability in pitch present in the pop texture. Output variability is important for the realism of generated textures, and was observed on all datasets as can be heard in the examples.

\begin{figure}[htb]
\begin{minipage}[b]{1.0\linewidth}
  \centering
  \centerline{\includegraphics[width=0.98\textwidth]{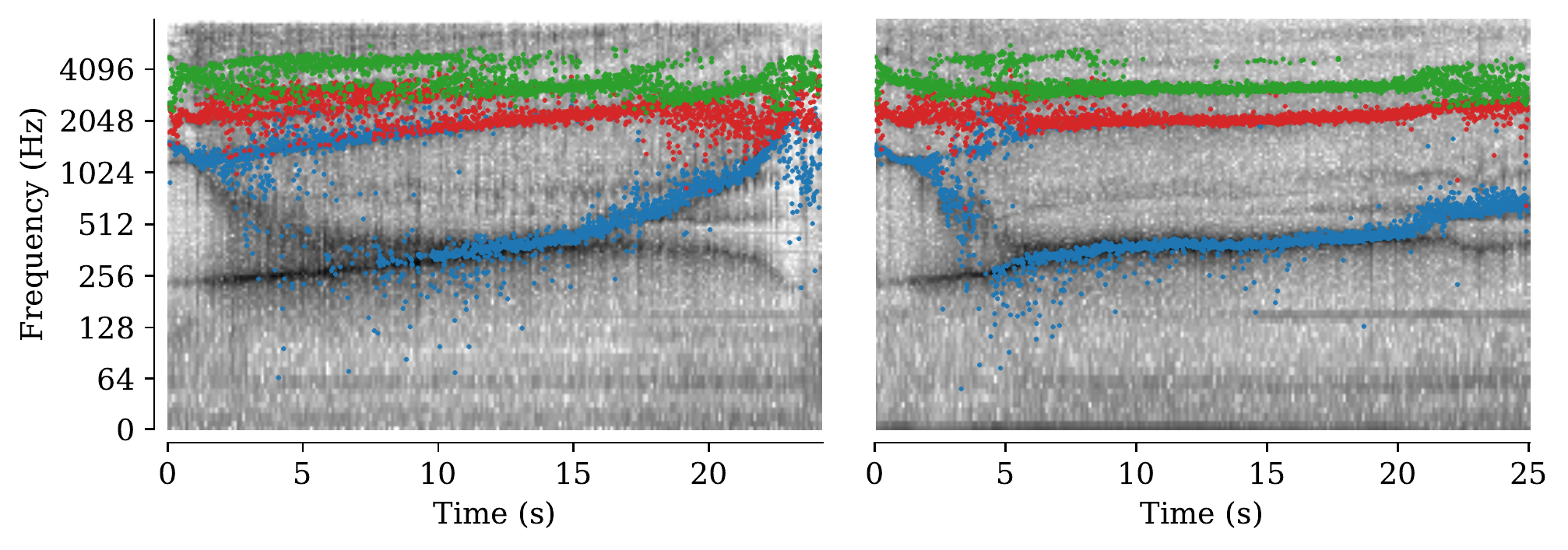}}
\end{minipage}
\caption{The 1st (blue), 2nd (red), and 3rd (green) formants overlaid on a spectrogram of the ``container filling'' texture for both real (left) and synthesised (right) data.
}
\label{fig:formant}
\end{figure}

The chromagram is less informative for unpitched or more stochastic textures that may contain multiple sound sources. For the ``container filling'' dataset of water textures examplifying this type of sound, we found formant analysis to be more informative. Formants are prominent bands of frequency with a concentration of energy resulting from acoustic resonances. The first three formants were extracted from both a real example from the test set and a synthesised sequence, as the fill parameter increased linearly from 0 to 1. The model was able to reproduce the individual formants at the correct frequency bands (Fig.\ref{fig:formant}). The formants in the synthesised examples were however flatter over time than the original, indicating that the overall perceived pitch did not change as much given the fill parameter. They were also not as well modelled during the transient sections at the start and end of the clip.

\subsection{Quantitative evaluation}
Correspondence between objective measures with perceptual quality have proven notoriously difficult to achieve. Several works in the field rely on metrics based upon deep learning-derived statistics found to correlate well with human evaluation. We evaluated the MTCRNN with different conditions against Wavenet on the ``container filling'' and ``pop'' datasets through their Fr{\'e}chet audio distance (FAD) \cite{kilgour2018fr}. The FAD compares statistics from real and generated data calculated from the embedding layer of a VGGish model pretrained on AudioSet. A lower FAD score denotes a smaller distance between the real and generated distributions in the embedding space of the VGGish model. The model variants tested include the full MTCRNN model which generates its own conditional parameters, a 1-tier model conditioned on real test set parameters, a 1-tier model conditioned directly on the control parameter, and their Wavenet equivalents (Table \ref{table:fad}). 1-tier here refers to the audio generating tier of the MTCRNN model absent of the higher level parameter predicting tiers. We note that the FAD scores do not reflect the performance with respect to capturing time dependencies beyond the time window of a frame, such as those present in the ``pop'' dataset. Both the MTCRNN and Wavenet conditioned only on the control parameter deviated from the expected timing of the pops, belying their relatively well performing FAD scores. The timing structure present in the original signals was far better captured with the inclusion of intermediate conditional parameters for either model type.

\begin{table}[tb]
\centering
\scriptsize
\begin{tabular}{@{}ll@{}}
	\toprule
	\multicolumn{1}{l}{\textbf{Model}} & \multicolumn{1}{l}{\textbf{FAD}} \\
	\midrule
	\multicolumn{2}{@{}l}{\textbf{``container filling'' dataset}} \\
    3-tier MTCRNN + generated cond. & 2.89 \\
    1-tier MTCRNN + real MFCC cond. & 0.617 \\
    1-tier MTCRNN + ``fill'' cond. & 8.59 \\
    Wavenet + real MFCC cond. & 1.76 \\
    Wavenet + ``fill'' cond. & 1.79 \\
    \midrule
	\multicolumn{2}{@{}l}{\textbf{``pop'' dataset}} \\
    2-tier MTCRNN + generated cond. & 1.53 \\
    1-tier MTCRNN + real onset strength \& RMSE cond. & 1.07 \\
    1-tier MTCRNN + ``rate'' cond. & 1.27 \\
    Wavenet + real onset strength \& RMSE cond. & 1.39 \\
    Wavenet + ``rate'' cond. & 1.17 \\
	\bottomrule
\end{tabular}
\caption{FAD scores for various datasets and models (lower better).}
\label{table:fad}
\end{table}

\begin{table}[tb]
\centering
\scriptsize
\begin{tabular}{@{}llll@{}}
	\toprule
	{\textbf{Dataset/Model}}  & \textbf{Value} & \textbf{Min} & \textbf{Max}\\
	\midrule
	Ground truth  & $0.045\pm0.043$  & 0.002 & 0.305\\
	3-tier MTCRNN  & $0.038\pm0.052$  & 0.002 & 0.298\\
	1-tier MTCRNN  & $0.295\pm0.144$  & 0.034 & 0.603\\
	\bottomrule
\end{tabular}
\caption{Spectral flatness of the results from a standard MTCRNN compared to a 1-tier variant and the ground truth (lower better).}
\label{table:spec_flatness}
\end{table}

The FAD scores showed that the MTCRNN performed comparably to Wavenet. With real parameter conditioning, the MTCRNN surpassed its Wavenet counterpart on the metric, suggesting that a relatively simple stacked RNN architecture is a viable alternative to dilated convolutions for textures, given the presence of supplementary conditional information. Despite the difference in the FAD, qualitative differences between the Wavenet and MTCRNN results were small. Training times were significantly shorter for the MTCRNN with fewer parameters for the independently trained tiers. When conditioned on generated parameters in the full model, the drop off in quality was not major, an arguably justifiable trade-off having gained the ability to generalise beyond the training data and synthesise arbitrary sequences. In contrast, models conditioned purely on the control parameter deteriorated the FAD, especially for the more complex ``container filling'' dataset. A comparison of the average spectral flatness (Table \ref{table:spec_flatness}), an indicator of how noise-like a sound is (a value close to 1 denotes a spectrum similar to white noise), confirms the higher noise floor present in the results of the 1-tier model. This further emphasises the importance of auxiliary information and justifies the multi-tier architecture.

\subsection{Exploration of parameter space}

The model was able to generalise beyond the discrete parameter values on which it was trained, mapping continuous control parameters to continuous acoustic characteristics (see links to sound examples in Table \ref{table:c7_data}). The flexibility afforded by the model with a comparatively sparse amount of data opens up many avenues for the creative use of the system. For instance, the model can be directed to synthesise sounds that are not physically possible in real life such as holding the ``container filling'' texture constant with an unchanging fill parameter, or even ``unfilling'' a container by reversing the direction of the fill parameter that was always increasing during the course of all training sets. Control parameters can be set to follow any arbitrary mathematical function. The engine dataset consists of steady-state examples at discrete ``rev'' parameter values controlling pitch and volume characteristics. A convincing accelerating/decelerating engine was generated by setting ``rev'' to a sine wave pattern. The model was also able maintain global parameters such as ``class'' in the heartbeat dataset while simultaneously accounting for a second, more dynamic ``rate'' parameter. Extrapolation beyond trained parameters was surprisingly good, at least for simpler sounds such as the ``pop'' texture. In tests, the correct underlying number of pops per unit time was maintained up to a ``rate'' of around 3, triple the maximum value of the parameter used in training.


\subsection{Style transfer}
Style transfer imposes certain properties of one domain, as captured by a model, onto another domain. For example, Timbretron \cite{huang2018timbretron} transfers timbre between musical pieces while keeping the musical note and rhythm structures intact. The MTCRNN can be adapted to transfer the longer range patterns captured by the upper tiers of one sound model to the sample-level modelling of another sound, using common conditioning parameters between them as an interface. For instance, RMSE and onset strength values synthesised by the ``pop'' model were provided as conditioning to the bottom tier of the ``geiger counter'' model to produce geiger clicks regularly spaced in time, even though such time patterns were never exposed to the audio model during training. Other applications include extending control parameters beyond the trained range the model, creating novel timbres, and accommodating new control parameters. Sound examples for these style transfer scenarios may be auditioned (\url{https://bit.ly/34fsr5A}).


\section{Conclusion}
\label{sec:conclusion}

We showcased the modelling of audio textures in the raw audio domain using the MTCRNN. The model learns acoustic characteristics across multiple timescales and connects its various components through conditioning without the need to explicitly propagate gradients through the entire network. Its hierarchical nature accounts for both local information like those pertaining to timbre or salient audio events in time, and slower changing characteristics such as pitch trajectories. This also enables a denser set of physically-relevant but less intuitive parameters used to support the synthesis (e.g. MFCCs, RMSE, onset strength) to be connected to simpler control parameters more directly relevant to the everyday experience of the user (e.g. rate, rev, fill). While establishing an optimal parameter space for textures analogous to MIDI was not the main focus of this work, and may be revisited in the future, having the flexibility to design control parameters without radically changing the model allows for the creation of sound models tailor-made for specific applications, especially interactive ones like video games or live music.

\bibliographystyle{IEEEbib}
\bibliography{refs}

\end{document}